\documentstyle[prl,aps,preprint]{revtex}

\begin{document}
\draft
\title{Charging effects in ultrasmall
quantum dots in the presence of time-varying fields}
\author{C. Bruder and H. Schoeller}
\address{
Theoretische Festk\"orperphysik, Universit\"at Karlsruhe, 76128 Karlsruhe,
Germany
}
\date{\today}
\maketitle
\begin{abstract}

The influence of charging effects on time-dependent transport
in small semiconductor quantum dots with arbitrary level spectra is studied.
Starting from an explicit time-dependent tunneling Hamiltonian, a
non-Markovian Master equation is derived which is also valid in the
nonlinear response regime. The many-body nonequilibrium distribution
functions of the dot are calculated and the I-V characteristic of the
structure including the displacement currents is obtained. New resonant
features show up in the Coulomb oscillations and in the Coulomb staircase, and
a new possibility to realize electronic pumps is described.
\end{abstract}

\narrowtext
The influence of the Coulomb interaction on low-temperature quantum
transport through metallic or semiconducting islands (quantum dots) has been
the subject of many theoretical and experimental papers (see, e.g.,
\cite{gd} and references therein). It shows up in a variety of
interesting effects like, e.g., Coulomb blockade and resonant tunneling
phenomena. However, the investigation of time-dependent perturbations
and of fluctuations of the electrodynamic environement has started only
recently, either neglecting charging effects
\cite{hns,johansson90,ct,liu,runge,pastawski2,fd,buett,mjw,da} or considering
the metallic case for time translational invariant systems
\cite{in,ofs}.
An explicit time dependence of the applied voltages introduces a new energy
scale $\hbar\omega$ in the problem, and a multitude of new effects is
expected to occur, some of them relevant to device applications such as
high-frequency oscillators.

In this letter, we consider an interacting quantum dot with an arbitrary
level spectrum subject to an explicit time-dependent classical field.
It can describe a periodic modulation of the Fermi energy in the leads
(i.e. time-dependent bias voltages) or time-dependent perturbations for
the quantum states in the dot. We are especially interested in the effects
of the Coulomb interaction in the limit of low tunneling rates but
finite level spacing, so that resonant tunneling
with thermally broadened line shapes will occur. In contrast to
rfs.~\cite{in,ofs}, we calculate the complete time-dependent (many-body)
distribution function of the dot using a non-Markovian Master equation
approach. As a consequence we obtain new features in the Coulomb oscillations
which can not be seen in the time-independent case or for continuous level
spectra like in metals. Furthermore, for certain asymmetric level structures
in the nonlinear response regime, a new mechanism to realize
an electronic pump is described.

An important consistency check for our formalism is the fact that the sum of
all currents into the system (including the displacement currents) is conserved
\cite{buett}. We have included the displacement currents
in our formalism within the Coulomb-blockade model to make our calculations
more realistic. However, since we are mostly interested in the case of low
capacitances (where charging effects are important), we do not obtain a
significant influence of the displacement currents on the I-V characteristics.

As a model for an interacting quantum dot coupled to two reservoirs by
tunnel junctions with capacitances $C_L$ and $C_R$, we will use the
time-dependent tunneling Hamiltonian
$H(t)=H_0(t)+V(t)+H_T$. Here,
\begin{equation}
H_0(t) = \sum_{\alpha=L,R} \sum_k \epsilon_{k\alpha} (t) a^\dagger_{k\alpha}
a_{k\alpha} + \sum_l \epsilon_l c^\dagger_l c_l
\label{1}
\end{equation}
describes non-interacting electrons in the reservoirs $\{ \alpha\}$
and in the dot.
A time-dependent shift of the Fermi energy of the electrons in the leads
has been included in $\epsilon_{k\alpha} (t)=\epsilon_{k\alpha}^0 +
\Delta_\alpha(t)$.
The tunneling part is given by $H_T=\sum_{\alpha}\sum_{kl} (T^\alpha_{kl}
a_{k\alpha}^\dagger c_l + c.c.)$, where $T^\alpha_{kl}$ denotes the tunneling
matrix element, and the Coulomb interaction is approximated by the
Coulomb-blockade model
\cite{Kul,Ave,GlMa}

\begin{equation}
V(t)=E_C (\hat N_D+n_0(t))^2
\label{2}
\end{equation}
corresponding to the equivalent circuit (SET-transistor) schematically drawn
in Fig. \ref{fig1}.
The particle number in the dot is denoted by $\hat N_D=\sum_l c_l^\dagger c_l$,
$E_C=e^2/2C$ is the charging energy with $ C=C_L+C_R+C_g$ and $en_0(t)=
C_LV_L(t) + C_RV_R(t)+C_gV_g(t)$ accounts for the time-dependent voltages of
the left and right reservoirs $eV_{L/R}(t) = \mu_{L/R}+\Delta_{L/R}(t)$
as well as a time dependent gate voltage $eV_g(t) = \mu_g+\Delta_g(t)$
applied to the quantum dot by the capacitance $C_g$.

The tunneling currents from the left/right reservoir into the quantum dot
are given by
\begin{equation}
I^{\text{Tun}}_{L/R}(t)=e {d \over dt} \left < \hat N_{L/R} \right > _
{\hat\rho(t)}
\label{3}
\end{equation}
where $\hat N_\alpha = \sum_k a^\dagger_{k\alpha} a_{k\alpha} $ is the particle
number in reservoir $\alpha$ and $ \hat\rho(t)$ denotes the density matrix
of the system at time $t$.
The expectation value $<\hat N_\alpha>$ can only change due to
tunneling processes, therefore, the capacitive currents $\dot Q_{L/R}$
($ Q_{L/R}$ is the charge on the left/right tunnel junction, see Fig.
\ref{fig1}) are not included in (\ref{3}) and have to be added separately.
The displacement currents are fixed by electrostatic considerations.
The total current is
\begin{eqnarray}
I_L=&&{C_R C_L \over e C}(\dot \Delta_L-\dot \Delta_R)
+{C_g C_L \over e C}(\dot \Delta_L-\dot \Delta_g)\nonumber\\
&&+{C_R +C_g \over e C}I^{\text{Tun}}_{L}
-{C_L \over e C}I^{\text{Tun}}_{R}
\label{4}
\end{eqnarray}
and $I_R$ is obtained by interchanging $L$ and $R$. The capacitive
current in the gate electrode is denoted by $I_g$. A direct consequence is
the conservation of all currents $I_R+I_L+I_g=0$, as emphasized in recent
works by B\"uttiker et al \cite{buett}.
For $C_g=0$, we obtain at each $t$ $I_L(t)=-I_R(t)$, but the tunneling
currents (\ref{3})
do not have such a property since the charge $Q(t)$ on the island is
a function of time \cite{rem1}. Generally, Eq. (\ref{3}) gives the
total current only if it is used to calculate the time-averaged current
\cite{mjw}. If the capacitances are large,
the first two terms on the r.h.s. of (\ref{4}) can be more important than the
tunneling part. For low capacitances, on the other hand, the
Coulomb energy is high and charging effects like the Coulomb-blockade
will appear. We are interested in this limit and we will have to
include the interaction (\ref{2}) in a nonperturbative way.

In order to calculate the tunneling currents (\ref{3}), we perform a
time-dependent unitary transformation defined by
$\bar{H}(t)=U(t)H(t)U^{-1}(t) + i\hbar \dot U(t)U^{-1}(t)$ with
\begin{equation}
U(t)=\exp({i \over \hbar}\int_{-\infty}^t d\tau
[\sum_\alpha \Delta_\alpha(\tau) \hat N_\alpha +\Delta_D(\tau)\hat N_D])
\label{5}
\end{equation}
where $\Delta_D(t)=C^{-1}\sum_{i=L,R,g}C_i\Delta_i(t)$ describes an
effective time-dependent perturbation on the dot. The result is
$\bar{H}(t)=\bar{H}_0+\bar{V}+\bar{H}_T(t)$
where $\bar{H}_0$ is the time-independent part of
(\ref{1}) (a constant shift of all
energies has been included in $\epsilon_l$), $\bar{V} =E_C \hat N_D
(\hat N_D-1)$ and
$\bar{H}_T(t)$ contains now time dependent tunneling matrix elements
$
\bar{T}_{kl}^\alpha(t)=T_{kl}^\alpha \exp({i \over \hbar}\int_{-\infty}^t d\tau
[\Delta_\alpha(\tau)-\Delta_D(\tau)]),
$
and for an
arbitrary interaction $\bar{V}$ on the dot the current $I_{L/R}(t)$
will depend only on the difference of the external voltages
$\Delta_{L/R}(t)-\Delta_g(t)$. The differences that appear in the
Hamiltonian, $\Delta_\alpha-\Delta_D$, are linear functionals of
$\Delta_\alpha-\Delta_g$, vic.
$\Delta_L-\Delta_D=[(C_R+C_g)(\Delta_L-\Delta_g)-C_R(\Delta_R-\Delta_g)]/C$,
and analogously for $\Delta_R-\Delta_g$. Eq. (\ref{3}) is unchanged by the
unitary transformation since $\hat N_{L/R}$ commutes with $U(t)$. A systematic
perturbation expansion in $\bar{H}_T$ then yields
\begin{equation}
I^{\text{Tun}}_\alpha(t)=e\sum_{rr'}\int_{-\infty}^t dt' \Gamma_{rr'}(t,t')
[N_\alpha(r')-N_\alpha(r)]P_r(t')
\label{7}
\end{equation}
where $|r>$ are the eigenstates of $\bar{H}_0+\bar{V}$ with energy $E_r$,
$P_r(t)=<r|\hat\rho(t)|r>$ are the diagonal elements of the density matrix,
$N_\alpha(r)=<r|\hat N_\alpha |r>$ is the particle number in reservoir
$\alpha$ for the state $|r>$, and $\Gamma_{rr'}(t,t')$ are the
time-dependent matrix elements
\begin{eqnarray}
\Gamma_{rr'}(t,t')&=&{2 \over \hbar^2} Re \{ <r|\bar{H}_T(t)|r'>\nonumber\\
& &\hspace{-1cm}
\times <r'|\bar{H}_T(t')|r> \exp({i \over \hbar}(E_r-E_{r'}+i\frac{\Gamma}{2})
(t-t'))\}
\label{8}
\end{eqnarray}
for the transition $r'$ (at time $t'$) $\to r$ (at time $t$).
Non-diagonal elements of $\hat\rho(t)$ are of higher order in $\bar{H}_T$
compared with the diagonal part $P_r(t)$ and do not contribute to
the lowest order result (\ref{7}). This is equivalent to neglecting
initial correlations between the dot and the reservoirs. Furthermore,
we have included a finite lifetime $\tau=\frac{2}{\Gamma}$ in (\ref{8})
due to tunneling, where $\Gamma=\sum_\alpha\Gamma_\alpha$ and
$\Gamma_\alpha=2\pi\sum_k|T_{kl}^\alpha|^2\delta(E-\epsilon_{k\alpha})$ is the
tunneling rate into reservoir $\alpha$ (for simplicity, we assume
$\Gamma_\alpha$ to be independent of the state $l$ and the energy).
This is equivalent to the well-known inclusion of collisional broadening
effects in kinetic equations \cite{sch92,wil87}.

In the same way, we expand $\dot P_r(t)$ in powers of $\bar{H}_T$ and obtain
the following non-Markovian master equation
\begin{equation}
\dot{P}_r(t)=\sum_{r'} \int_{-\infty}^t dt' \Gamma_{rr'}(t,t') [P_{r'}(t')
-P_r(t)].
\label{9}
\end{equation}

Eq.~(\ref{7}) and (\ref{9}) are the result of our work in general form. They
allow us to treat time-dependent, non-linear transport problems in the
Coulomb blockade regime.
Although they have been derived in the limit of low tunneling rates
($\Gamma\ll kT$), the inclusion of a finite lifetime in (\ref{8}) softens
this restriction and leads to the exact result for the noninteracting case.
In the presence of charging effects, it has recently been shown
both experimentally \cite{mei} and theoretically \cite{sch93}, that
$\frac{\Gamma}{2}$ describes the broadening of the
Coulomb oscillation peaks in good approximation for discrete level spectra
($\Delta E>\Gamma$) and temperatures ($kT \sim \Gamma$) that are not too low.
For time-independent perturbations in the Markov-limit, we recover
the well-known Pauli-Master-equation \cite{kam}.

We will now approximate the interaction by the Coulomb-blockade model
(\ref{2}), which has turned out to be sufficient to describe the charging
effects in small quantum dots qualitatively, provided that the number of
electrons is not too small
\cite{pwh}. We therefore expect that it is also
appropriate for the description of time-dependent phenomena. The
eigenstates of $\bar{H}_0+\bar{V}$ are then given trivially in the occupation
number representation.

For weak tunneling and large reservoirs, we can factorize $P_r(t)=P_\phi
^{eq}P_s(t)$ into an equilibrium part $P^{eq}_\phi$ for the reservoirs
(described by the chemical potentials $\mu_\alpha$ according to the
time-independent part $\bar{H}_0$) and a part $P_s(t)$ for the dot. In
order to find the stationary value for $P_s(t)$, we take a periodic
modulation $\Delta^\alpha (t)=\Delta^\alpha_0 \sin\omega t$ ($\Delta^\alpha
=\Delta_\alpha-\Delta_D$) and perform a Fourier transformation
$P_s(t)=\sum_{m=-\infty}^\infty \tilde{P}_s(m)\exp(-im\omega t)$ with the
result
\begin{eqnarray}
(N_s-i\frac{m\hbar\omega}{\Gamma})\tilde{P}_s(m)\,&=&\,\sum_ln_l(s_l)
\tilde{P}_{s_l}(m) \nonumber \\ && \hspace{-2cm}
+\left[\sum_n\sum_{\alpha l}\frac{\Gamma_\alpha}{\Gamma}\,F^\alpha_{nm}
(E^{\alpha l}_{N_s-n_l(s)})\,(2n_l(s)-1)(\tilde{P}_s(n)+\tilde{P}_{s_l}(n))
+(m\rightarrow -m)^*\right]
\label{10}
\end{eqnarray}
where $s=|\{n_l(s)\}\rangle$ denotes a state of the dot characterized
by the occupation numbers $n_l$ of its single particle levels $l$, $s_l$ is
the state obtained from $s$ by reversing the occupation of level $l$,
i.e., $n_l(s_l)=1-n_l(s)$, $N_s=\sum_ln_l(s)$ and $E^{\alpha l}_N=\epsilon_l
+U^+_N-\mu_\alpha+i\frac{\Gamma}{2}$ where $U^+_N=2NE_c$ is the
change of the Coulomb energy due to the addition of one particle to the dot.
Furthermore, with $\beta=1/kT$ and $f(E)=(\exp{\beta E}+1)^{-1}$,
we have defined
\begin{eqnarray}
F^\alpha_{nm}(E)&=&i^{m-n}\sum_k J_{k+n}(\frac{\Delta^\alpha_0}{\hbar\omega})
J_{k+m}(\frac{\Delta^\alpha_0}{\hbar\omega})
\,Y(E+k\hbar\omega) \label{11}\\
Y(E)&=&\frac{1}{2}f(E)+\frac{1}{4\pi i}\left[\psi(\frac{1}{2}+i
\frac{\beta E}{2\pi})
+\psi(\frac{1}{2}-i\frac{\beta E}{2\pi})\right] \label{12}
\end{eqnarray}
where $J_k$ and $\psi$ denote the Bessel and digamma functions. Finally,
the tunneling current reads
\begin{equation}
\tilde{I}^{Tun}_\alpha(m)=\frac{e}{\hbar}\Gamma_\alpha\left\{\sum_s N_s
P_s(m)-\left[\sum_n\sum_{sl}\,F^\alpha_{nm}(E^{\alpha l}_{N_s-n_l(s)})\,
\tilde{P}_s(n)+(m\rightarrow -m)^*\right]\right\}
\label{13}
\end{equation}
Eq.~(\ref{10}) is a linear set of equations for the Fourier components of
the probability function $P_s(t)$. It can be solved numerically
in a straightforward way and gives directly the time-dependence of all
many-body distribution functions of the dot together with the tunneling
currents (\ref{13}). It can be applied to the nonlinear response regime
(i.e. high bias voltages or high time-dependent perturbations), small or
high frequencies, arbitrarily strong Coulomb-interaction and discrete
level spectra. Furthermore, no assumptions are necessary for the
distribution functions of the dot which can differ considerably from the
equilibrium value. For the time-independent case, Eq.~(\ref{10}) reduces to
the Pauli-Master equation which has been used in
Refs.~\cite{beenakker91,hbs,akl,pwh} to describe Coulomb oscillations and
the Coulomb staircase
in small quantum dots. The effect of time-dependent perturbations show up
clearly in the argument $E+k\hbar\omega$ (with $E=E^{\alpha l}_N$) in
(\ref{11}). It coincides with the one-particle excitation (tunneling)
spectrum $\epsilon_l+U^+_N+k\hbar\omega$ of the dot since the electrons can
now absorb or emit energy quanta $\hbar\omega$. For the special case of
one state in the dot, the solution of (\ref{10}) and (\ref{13}) reads
\begin{equation}
\tilde{I}_\alpha^{Tun}(m)=\frac{e}{\hbar}\Gamma_\alpha \left(\frac{\sum_\beta
\Gamma_\beta A_{\beta m}}{\Gamma-im\hbar\omega}-A_{\alpha m}\right)
\label{14}
\end{equation}
where $A_{\alpha m}=F^\alpha_{0m}(E^\alpha_0)+F^\alpha_{0,-m}(E^\alpha_0)^*$.
It is identical to the result obtained in Ref.~\cite{mjw} where
interesting current oscillations have been derived from (\ref{14}). Using
our approach, we have studied this effect for many levels with
Coulomb-interaction and have seen that it still remains
observable for discrete level spectra
($\Delta E>$ bias voltage) or, for high charging energy, even
in the case of continuous spectra like in metals. The capacitive currents
will not change this picture significantly, since they are small for the
low capacitances considered here.

We will now discuss several applications of the formalism presented above.
As a simple example we consider the DC-component of the current through
a double-barrier quantum well with two degenerate states ($\epsilon_1=
\epsilon_2=\epsilon$). Applying an AC gate voltage with frequency $\omega$
to the dot ($\Delta_L=\Delta_R=0$, $\Delta_g\neq 0$), the Coulomb staircase
reveals a fine-structure (Fig.~2) with additional steps at $\epsilon+
2NE_c+k\hbar\omega$ ($k\neq 0$ ; $N=0,1$) due to the new excitations from
the time-dependent perturbation.
Even in the case of many levels in the dot
with a continuous spectrum (where no fine structure due to the finite
level spacing can be observed), these side steps will remain. In the same
way, also the Coulomb oscillations will show additional resonances at a
distance $n\hbar\omega$ ($n=0,\pm 1,\pm 2,\ldots$) from the conventional main
peaks at $\epsilon+2NE_c$ ($N=0,1$). However, the situation changes
drastically if the two single particle levels are no longer degenerate
($\epsilon_1<\epsilon_2$). In the static case, we get two Coulomb peaks
at $\epsilon_1$ and $\epsilon_2+2E_c$ (inset Fig.~2, dashed curve), i.e.
the excitations $\epsilon_2$ and $\epsilon_1+2E_c$ are hidden without time
dependent perturbations. The reason is that on lowering the gate voltage
(for small bias voltage and particle number $N=0$), the first excitation
will be $N=0\rightarrow N=1$, occupying the level $l=1$, i.e., the first
peak is at $\epsilon_1$. After this transition, the dot contains exactly
one particle which can not escape due to the Coulomb blockade.
The next possible excitation for $N=1\rightarrow N=2$ is at energy
$\epsilon_2+2E_c$ and the ones with $\epsilon_2$ and $\epsilon_1+2E_c$ are
not visible. In the time-dependent case, the particle in the dot can
absorb modulation quanta of energy $\hbar\omega$ and thus it can leave the
dot or occupy the level $l=2$. Therefore, the complete tunneling
spectrum will show up in the Coulomb oscillations (see inset Fig.~2) in
contrast to the static case. In other words, the effect
of a finite charging energy in the presence of time dependent perturbations
is not only a simple shift of the resonances at $\epsilon_l$ ($l=1,2,\ldots$)
by $(l-1)2E_c$ but instead a splitting into many peaks at $\epsilon_l
+2NE_c$ ($N=0,1,2,\ldots$) where the amplitude is maximal at $N=l-1$ and the
peaks with $N=l-1+k$ correspond to absorption (emission) of energy
quanta $k\hbar\omega$. As in the case with the Coulomb staircase,
these fine structures are not observable for a continuous spectrum in the dot
where only the satellite peaks shifted by $k\hbar\omega$ will remain.

Another possibility is to apply an AC-signal to the reservoirs
such that
($\Delta_L-\Delta_D\neq 0,\Delta_R-\Delta_D=0$). Even for zero DC-bias voltage,
we obtain a positive or negative DC-current through the quantum well due
to the absorption or emission of energy quanta in the left reservoir.
The effect is possible only if there is an asymmetry of level(s) in the dot
with respect to the chemical potential ($\mu=\mu_L=\mu_R$) of the reservoirs:
if the level closest to $\mu$ is above $\mu$, there will be a positive
current, and vice versa.
Consequently, the system acts like an electron pump in this case and the
form of the Coulomb-oscillation peaks are changed dramatically as
illustrated in Fig. {\ref{fig4}. This example of a non-linear
response is a generic result as can be seen by taking the ($m=0$)-component of
(\ref{14}) for the case of a single nondegenerate level in the dot and
expanding to second order in $\Delta^\alpha$. For $\omega \to 0$, we obtain
$I_L^{DC} \sim f''(\epsilon_0-\mu)(\Delta^L)^2$.
The second derivative of the Fermi function will produce the positive
and negative regions seen in Fig. \ref{fig4}.

In conclusion, we have presented a formalism that allows us to calculate
the frequency-dependent $I-V$-characteristics of strongly interacting quantum
dots. It takes into account non-equilibrium many body distribution
functions of the dot, it is not restricted to the linear response regime, and
it is exact in the limit of small tunneling rates.
We have also included the contribution of the displacement currents
caused by the capacities and gate electrodes. The presence of five
energy scales leads to a multitude of physical situations that will
be studied in future work.

\acknowledgments
It is a pleasure to thank G. Sch\"on for initiating this work and for many
helpful discussions. We would also like to acknowledge stimulating discussions
with A. Groshev, F. Hekking, Y.V. Nazarov, A. Odintsov, and especially
with M. B\"uttiker. This work was supported by the Swiss National Science
Foundation (H.S.)and by the 'Deutsche Forschungsgemeinschaft' as part of
'Sonderforschungsbereich 195'.


\begin{references}
\bibitem{rem1} For $C_L=C_R+C_g$ the tunneling part of (\ref{4})
reduces to $(I_L^{\text{Tun}}-I_R^{\text{Tun}})/2$, an expression sometimes
referred to as the Ramo-Shockley theorem \cite{ct}. Note that it neglects
the capacitive contributions.
\end{references}

\begin{figure}
\caption{Equivalent circuit of the double-barrier structure in the static
case. $C_L, C_R$ are the capacitances between the dot and the leads,
and $C_g$ is the capacitance between the dot and the gate electrode.
$Q$ is the excess charge on the dot, and the $Q_i$ are the charges at
the capacitor plates.
}
\label{fig1}
\end{figure}

\begin{figure}
\caption{DC current versus $eV_{\text{Bias}}$ for
$T=5$, $E_C=75$, $\omega=50$,
$\Delta^L=\Delta^R=0$ (dashed line),
$\Delta^L=\Delta^R=50$ (solid line),
$C_L=C_R=10^{-4}$.
The Coulomb staircase is modified by additional steps.
Energies are measured in units of the total tunneling rate,
$\Gamma=\Gamma_L+\Gamma_R$ where $\Gamma_L=\Gamma_R$,
currents in units of $e\Gamma/\hbar$, frequencies in units of
$\Gamma/\hbar$, and capacities in units of $e^2/\Gamma$.
Inset: DC current versus $eV_g C/C_g$ for two nondegenerate levels
with $\epsilon_{1,2}=\mp 12.5$, other parameters as before. The Coulomb peaks
are seen to develop side bands shifted
by multiples of $\hbar\omega$. In addition, there are peaks corresponding
to the single particle levels in the dot.
}
\label{fig2}
\end{figure}

\begin{figure}
\caption{DC current versus $eV_g C/C_g$ at zero DC bias voltage and
$\Delta^L=30$, $\Delta^R=0$. Here, $T=2$,
$\omega=20$, and the other parameters are as before.
The structure acts as
an electron pump. The direction of the current depends on the position
of the two many-particle levels with respect to the chemical potentials
of the left and right leads.
}
\label{fig4}
\end{figure}

\end{document}